\documentclass[preprint,showkeys,prb,amsmath,amssymb]{revtex4}

\usepackage{graphicx}
\usepackage{dcolumn}

\begin{document}


\title{Early stage morphology of quench condensed Ag, Pb and Pb/Ag hybrid films}

\author{Zhenyi Long}
\author{James M. Valles, Jr.}
 \email{valles@physics.brown.edu}
\affiliation{Department of Physics, Brown University, Providence, RI 02912}


\date{\today}

\begin{abstract}
{\it In situ} Scanning Tunneling Microscopy (STM) has been used to study the morphology of Ag, Pb and Pb/Ag bilayer films fabricated by quench condensation of the elements onto cold (T=77K), inert and atomically flat Highly Oriented Pyrolytic Graphite (HOPG) substrates. All films are thinner than 10 nm and show a granular structure that is consistent with earlier studies of QC films. The average lateral diameter, $\overline {2r}$, of the Ag grains, however, depends on whether the Ag is deposited directly on HOPG ($\overline {2r}$  = 13 nm) or on a Pb film consisting of a single layer of Pb grains ($\overline {2r}$ = 26.8 nm). In addition, the critical thickness for electrical conduction ($d_{G}$) of Pb/Ag films on inert glass substrates is substantially larger than for pure Ag films. These results are evidence that the structure of the underlying substrate exerts an influence on the size of the grains in QC films. We propose a qualitative explanation for this previously unencountered phenomenon.  

\end{abstract}


\maketitle

Metal films quench condensed from vapor onto substrates held at temperatures well below the metal melting temperature become electrically continuous even when very thin. This attribute has made quench condensed (QC) films useful for numerous studies of superconductivity, electron localization, and quantum phase transitions in two dimensions \cite{Buckel, Buckel2, Bulow, Strongin, Bergmann, Hsu, Liu, Dynes, Haviland, White, Jaeger, Barber, Markiewicz}. In addition, the growth mode of QC films has garnered attention since the familiar, thermally activated processes that dominate film structure formation at high temperatures cannot occur at $T_s < 0.1 T_M$, where $T_s$ and $T_M$ are the substrate and metal melting temperatures, respectively. QC film structures have been divided into two categories that depend on the substrate characteristics. QC films deposited on an inert substrate such as glass or Highly Oriented Pyrolytic Graphite (HOPG) are believed to form in a granular structure. QC films deposited on a predeposited Ge layer or similarly strongly interacting substrate are believed to form in a homogenous, amorphous structure. In this paper, we consider the category of granular films and present results that provide new insight into the factors governing the formation and geometry of their grains.  

Originally, the existence of granular structure was inferred from electrical transport measurements. Granular films became electrically conducting at a mass equivalent thickness, $d=d_G$, that corresponds to 10 or more monolayers. In the vicinity of $d_G$, the conductance increases nearly exponentially with $d$. This behavior led to the proposal that three dimensional islands or grains, separated by vacuum gaps, were forming on the substrate. The exponential dependence was consistent with electron tunneling dominating the interisland transport\cite{Strongin}. Danilov and coworkers \cite{Danilov, Danilov2} challenged this idea, doubting that islands could form in the absence of thermally activated processes. Instead, they proposed that for $d<d_G$ the metal atoms formed a novel, low density, disordered, insulating phase. For $d>d_G$ this phase became unstable to the formation of metallic crystalline inclusions. Electron tunneling between these inclusions and through the insulating phase dominated the transport in this regime.  

	Recent {\it{in situ}} low temperature STM studies on granular QC Au and Pb films \cite{Ekinci1998, Ekinci1999}  revealed that the evolution of their structure is best described by a combination of these two models. Films with $d<<d_G$ form an amorphous, non-conducting structure up to a critical thickness $d_c$. Just beyond $d_c$, a monolayer of crystalline, platelike grains with lateral diameter to height ratios of about 5, formed out of this amorphous structure, giving the films an islanded appearance. Subsequent depositions led to the formation of similar grains atop the first monolayer. These upper grains bridged the lower grains to provide the coupling necessary for electrical conduction. It was conjectured that the grain formation out of the amorphous phase was driven by the heat released through the crystallization process. Thus, the size and shape of the grains was determined by the heat of crystallization and the surface energies of different crystalline orientations. Since grains in the upper layers were similar to the first layer of grains it was concluded that similar processes drove their formation.
  
	The work presented in this paper provides evidence that the growth mode described above needs to be modified to correctly describe the formation of the upper layer of grains. It was originally motivated by the need to understand the structure of ultrathin, granular bilayer films of Pb and Ag employed in investigations of the superconducting proximity effect \cite{Merchant,Kouh,Long}. We performed {\it{in-situ}} STM studies and transport measurements on a series of QC granular Ag, Pb and Ag on Pb (Pb/Ag) hybrid films with thicknesses less than 10 nm. While we observed that QC Ag, Pb and Pb/Ag hybrid films on HOPG substrates exhibit the typical granular structure, we found that the average size of the Ag grains depended on whether they formed directly on the HOPG or on top of a layer of Pb grains. Ag grains on Pb assume nearly the Pb grain size and shape. They are a factor of two larger than Ag grains formed directly on HOPG. In addition, $d_G$ of the Pb/Ag hybrid films is greater than $d_G$ of the pure Ag film by more than a factor of two. These results suggest that the structure in the substrate exerts a strong influence on the structure of QC films. We propose modifications to the film growth model described above to account for these observations.  
 
The STM studies were carried out in a low-temperature STM cryostat equipped with multiple thermal evaporation sources. A more detailed description of the system can be found elsewhere \cite{Ekinci1997}. We used a thin slab of Highly Oriented Pyrolytic Graphite (HOPG) as the substrate. The HOPG substrate has many advantages that make it an ideal substrate. It is conductive, which makes tunneling into discontinuous films possible. It can be cleaved to produce a fresh substrate for each deposition. Furthermore, because it is inert, QC granular films on HOPG possess the same morphology as films deposited on other inert substrates such as glass \cite{Ekinci1998, Ekinci1999}. After mounting the substrate and tip, we evacuated the experimental chamber to $5 \times 10^{-7}$ Torr with a turbo pump. Then the whole chamber was immersed in liquid nitrogen. The substrate was held at 77K throughout the evaporations and STM imaging. A quartz crystal microbalance was employed to monitor the film thickness. With this apparatus, we were able to fabricate a series of films and monitor their morphology without breaking vacuum or warming. To fabricate the Pb/Ag hybrid films, we deposited the Pb film first with a thickness $d_{Pb} \sim$ 2.8 nm, which produces a nearly complete single layer of Pb grains \cite{Ekinci1999}. We then evaporated Ag on top of the Pb film. STM images were acquired no more than 10 mins after the evaporation.
 
The pair of elements, Pb and Ag, were chosen for the proximity effect studies because their equilibrium phase diagram indicates that they not form alloys \cite{Merchant,Kouh,Long}. Thus, the interfaces between the superconductor and normal metal grains could be presumed sharp. This presumption is central to our interpretation and discussion of the STM results presented here. That is, we take the imaged grains to be either pure Pb or pure Ag.

The transport experiments were performed in a second $^4$He pot cryostat. That cryostat was immersed in liquid Helium and the films were deposited onto fire polished glass substrates held at 8K. Au/Ge contact pads were deposited on the substrate prior to cryostat mounting. QC Ag, Pb and Pb/Ag hybrid films were fabricated and measured {\it in situ} in this UHV environment. Previous studies have shown that the morphology of QC Pb and Au films deposited on substrates of T=77K and T=4K do not show any noticeable differences \cite{EkinciThesis, Ekinci19982}. Thus, we assume that the morphology of the Ag and Pb/Ag films deposited on a T=8K substrate is the same as the corresponding QC film deposited at 77K. We used standard 4-terminal DC techniques to measure the film resistances.

\begin{figure}
\includegraphics{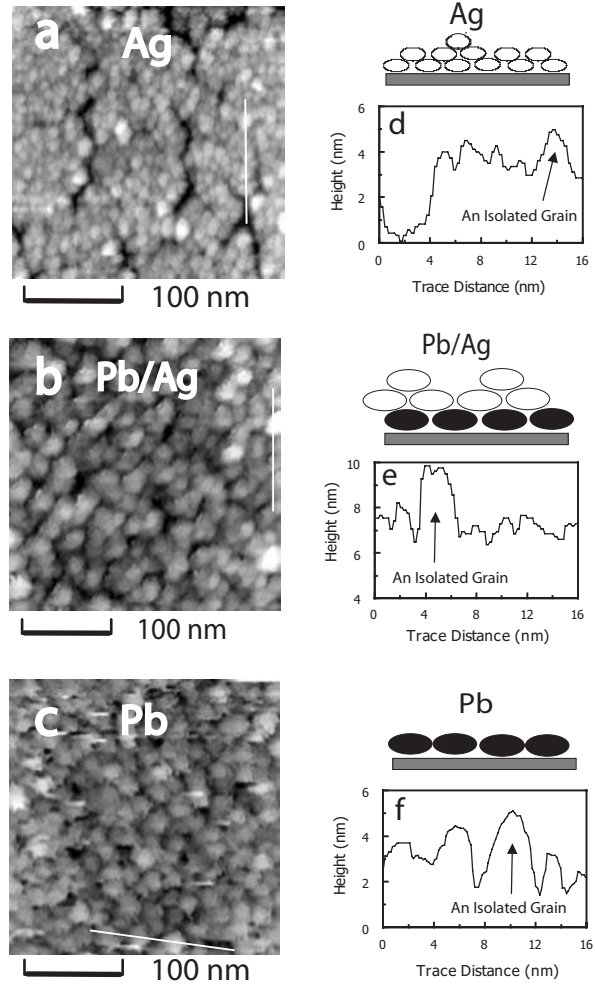}
\caption{{\it In situ} STM images of QC Ag, Pb/Ag, and Pb films on the same scan areas 346 $\times$ 346 nm. (a)A Ag film with $d_{Ag}$ = 3.0 nm. (b)A Pb/Ag film with $d_{Ag}$ = 4.2 nm and $d_{Pb}$ = 2.8 nm. (c) A Pb film with $d_{Pb}$ = 2.8 nm.
The corresponding line scans are displayed in (d), (e), (f) respectively.}
\end{figure}

{\it{In situ}} STM images of pure Ag ($d_{Ag}$ = 3.0 nm), pure Pb ($d_{Pb}$ = 2.8 nm), and Pb/Ag ($d_{Pb}$ = 2.8 nm and $d_{Ag}$ = 4.2 nm) QC films are compared in Fig. 1. For convenience, the scan areas of the three images are all the same, 346 nm $\times$ 346 nm. All images show the typical granular structure of QC films on inert substrates. Qualitatively, it is clear that the average lateral diameter, $\overline {2r}$, of a grain in the pure Ag film is substantially smaller than that in the pure Pb film. On the other hand, $\overline {2r}$ for the Ag grains in the Pb/Ag hybrid film is very close to the pure Pb film grain size. Line scans of the images (Figs. 1d-f) reveal that the height of the Ag grains is also greater in the Pb/Ag hybrid film compared to a pure Ag film. Some isolated grains in the film are higher (whiter) than the others, suggesting that they belong to the next layer of grains. Consequently, a line scan across such an isolated grain allows a measurement of the grain height\cite{Ekinci1998, Ekinci1999}. The line scans in Figs. 1d-1f, roughly indicate that heights of the grains in the pure Ag film are approximately 2 nm and thus, roughly 1/2 as tall as those in the pure Pb and Pb/Ag hybrid films. In the case of the QC Ag film (Fig. 1a), cracks are randomly distributed in the film structure. Presuming that the crack extends to the HOPG surface, the height difference between the bottom of the crack and the edge measures the geometrical film thickness as 4 nm. The geometrical thickness differs from the nominal thickness (3 nm) measured with the quartz crystal microbalance. This difference can be easily attributed to a loose packing of the grains which makes the film mass density less than the individual grain mass density. 

\begin{figure}
\includegraphics{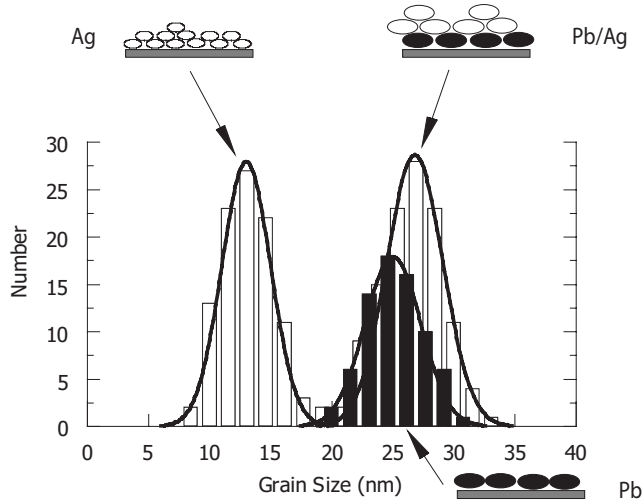}
\caption{Histograms of lateral grain diameters of the films in FIG. 1. The histograms can be roughly fitted by the normal distribution. The fitting parameters are: for the Ag film, $\overline {2r}$ = 13 nm, $\sigma_{r}$ = 2 nm; for the Pb/Ag film, $\overline {2r}$ = 26.8 nm, $\sigma_{r}$ = 2.3 nm; for the Pb film, $\overline {2r}$ = 25 nm, $\sigma_{r}$ = 2.3 nm}
\end{figure}

A more quantitative analysis of the grain sizes is illustrated in Fig. 2. Fits of circles to the edges of grains were used to obtain their lateral dimensions. Because most grains have a regular plate shape, the error of this kind of measurements is no more than 10 \%. Histograms of the grain diameters in each of the images in Fig. 1 are shown in Fig. 2. These histograms can be approximately described by the normal distribution, $N(2r)=N_{0} exp(-(2r-\overline{2r})^2/2\sigma_{r}^2))$, where $N_{0}$ is the total grain number, $\overline{2r}$ is the average grain diameter and $\sigma_{r}$ is the standard deviation for the grain diameter. The $\overline{2r}$ of the Ag grains in the hybrid film is 26.8 nm, which is almost two times larger than the $\overline{2r}$ = 13 nm for the grains in the Ag film. It is, however,very close to $\overline{2r}$ = 25 nm of the Pb grains. For all the films, $\sigma_{r}$ is no more than 15 \% of $\overline{2r}$. Furthermore, line scans (not included here) show that the grain heights of the Ag grains lie in the range from 1.8 to 2.4 nm while the grain heights of the Ag grains in a hybrid film and the Pb grains lie in the range from 3.3nm to 4.9 nm (see Table).

Measurements of the sheet resistance as a function of thickness were employed to determine the critical thicknesses for conduction, $d_G$, of QC Ag, Pb and Pb/Ag films. These are shown in Fig. 3 for pure Pb and Ag films and for a hybrid Pb/Ag film with $d_{Pb}$ = 2.96 nm. $d_G$ was defined as the thickness at which $R_N \sim 10^5 \Omega$ and its values are indicated by the arrows in Fig. 3. Because the sheet resistances drop steeply with increasing thickness at high sheet resistance, $d_G$ is relatively insensitive to the exact choice of a resistance criterion for conduction. Interestingly, $d_G$ for the pure Ag film is less than half the $d_G$ for the pure Pb and hybrid films (see Table~\ref{tab:table1}).  

\begin{figure}
\includegraphics{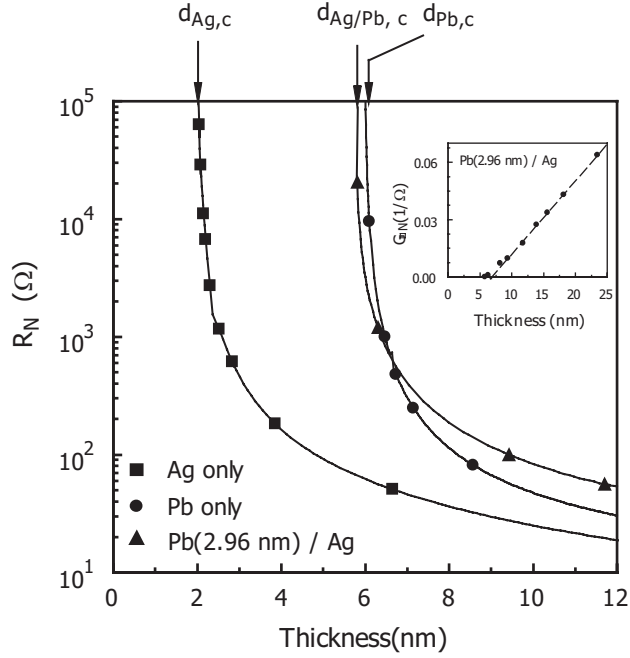}
\caption{Sheet resistance vs. thickness for Ag, Pb and Ag/Pb(2.96 nm) films. The critical thicknesses for electrical conduction are: $d_{Ag,c}$ = 2.0 nm, $d_{Ag/Pb,c}$ = 5.8 nm and $d_{Pb,c}$ = 6.0 nm. Inset: Sheet Conductance vs. thickness for the Pb(2.96 nm)/Ag hybrid film. The dashed line is the linear fit to the data.}
\end{figure}

These differences in $d_G$ can be semiquantitatively attributed to the difference in the Ag grain size in pure Ag and hybrid films. As we pointed out earlier, electrical continuity in granular films requires a second layer of grains. In the case of Pb films which have relatively large grains \cite{Ekinci1999}, the second layer must be nearly complete for conduction. While in the case of Au films which have grains comparable to pure Ag grains, the upper layer may only be about 30\% complete for the films to become continuous \cite{Ekinci1998}. Consequently, one would expect $d_G$ of pure Pb or Pb/Ag films to be about twice the average grain height, while $d_{G}$ of a pure Ag film to be about 1.3 times the average grain height. Combined with the fact that Ag grains in pure Ag films are about half as tall as in Pb/Ag films, one would anticipate the ratio of their $d_{G}$'s to be roughly 3. The experimental data in Fig. 3 and Table~\ref{tab:table1} show that $d_{Pb/Ag, G}/d_{Ag, G}$ is about 2.8, which is close to the expectation.  

\begin{table}
\caption{\label{tab:table1}A table of the grain dimension, critical thickness and mean free path of a few different materials}

\begin{ruledtabular}
\begin{tabular}{ccccc}
Material&$\overline{2r}$ (nm)&$h$ (nm)&$d_{G}$ (nm)&$l$ (nm)\\
\hline
Ag&13&$1.8 \sim 2.4$&2.1&3.75\\
Au\footnotemark[1]&15&$1.6 \sim 2.1$&2.2&\\
Pb&25&$3.3 \sim 4.9$ &6.0&2.3\\
Pb/Ag&26.8&$3.3 \sim 4.9$&5.8&3.3\\
\end{tabular}
\end{ruledtabular}
\footnotetext[1]{The data for Au are from Ekinci {\it{et al.}}'s work \cite{Ekinci1998}}
\end{table}

Beyond $d_{G}$, the sheet conductances of the films follow $G_{N} = \sigma d + G_{0}$ as illustrated by the example in the inset of Fig. 3. Here, $\sigma$ and $G_{0}$ are constants. This dependence suggests that the material for $d > d_{G}$ has homogeneous properties with a well defined conductivity, $\sigma$. In a free electron model, $\sigma$ is related to the mean free path $l$ through the formula $\sigma = ne^2l/v_{F}m$, where $n$ is the density of the electrons, $m$, $v_{F}$ are  the mass and the Fermi velocity of the electrons, respectively. The dashed line in the inset is a linear fit to $G_{N}$ vs. thickness for the hybrid film, Pb(2.96 nm)/Ag. By measuring the slope of the line, we obtain $\sigma = 4.52 \times 10^{6}$ 1/($\Omega\cdot$m). With $v_{F} = 1.39 \times 10^6$ m/s and $n = 5.85 \times 10^28$ 1/m$^3$ of Ag, and the mass and charge of the electron, we can derive that the mean free path $l$ = 3.3 nm. For these granular films, grain boundary scattering probably dominates their electron transport. So one can visualize that the grain dimensions, especially the grain height, defines the mean free path. The mean free paths of the three film systems considered here are comparable to the grain heights as shown in Table~\ref{tab:table1}.

To summarize the experiments, STM and transport measurements indicate that the size of Ag grains in QC films depends on whether they are deposited on a flat substrate or a granular film. The Ag grains are larger on Pb films and similar in size to the Pb grains. Furthermore, the transport experiments indicate that the size of the Ag grains does not seem to depend on how thick the Ag layer is on the Pb. If it did then the mean free path would be expected to change with thickness, which would lead to a deviation from linearity in the inset of Fig. 3.

The difference between the Ag grain size in pure Ag films and hybrid Pb/Ag films cannot be accounted for with the original model of QC granular film growth. In that model, deposited atoms first form an amorphous layer that, when thick enough, undergoes an amorphous-to-crystalline transition \cite{Ekinci19982}. The atoms initially just "stick" where they land because of the absence of thermal diffusion at low temperatures. As the film grows thicker and the influence of the substrate potential on the film structure wanes, however, the amorphous film becomes unstable.  It avalanches, driven by the heat of the crystallization, into a polycrystalline form . The resulting grain sizes and shapes are determined by the balance between surface and interfacial energies and the latent heat of crystallization\cite{Hu}. Within this model, the size and shape of Ag grains in QC films should not depend on whether the Ag is deposited on HOPG or granular Pb.   

Our observations suggest that the underlying layer on which a QC film is deposited serves as a template that exerts control on the subsequent film structure. Recent theories suggest how. As shown by the STM images, a QC Pb film presents a substrate with atomically flat grain surfaces over 10 nm distances and grain boundaries with a high concentration of step edges. It has been proposed that adatoms deposited on relatively cold substrates tend to accumulate at step edges \cite{Egelhoff, Ferris}. Furthermore, the atoms are believed to funnel down to lower sites of adsorption if they are deposited around step edges \cite{Evans, Stoldt}. These factors create a tendency for deposited atoms to accumulate more at the boundaries between grains rather than on the upper surfaces of grains. 

Thus we arrive at the following scenario for the growth of Ag grains on Pb. Ag grains first form in the intergrain regions where most atoms have accumulated. They grow outward by consuming atoms on the flat upper surfaces of the lower layer grains. The heat of crystallization provides the energy for atoms in the grain vicinity to move to its surface. As more Ag atoms are deposited, the grains continue growing until they impinge on a neighboring Ag grain. Neighboring grains are unlikely to coalesce at low temperatures because the energy barrier for grain coalescence is large. Consequently, the final size of a Ag grain is dictated by the spacing of nucleation sites on the substrate. Since this spacing is determined by the size of the Pb grains, the Ag grains assume the Pb grain size. As all of the grains impinge on neighbors, new grains nucleate at their grain boundaries and a new, upper layer of grains begins to form. 

In conclusion, we have observed that the size of grains in granular quench condensed films depend crucially on the substrate structure. Ag grains formed on top of Pb grains assume the same size as the Pb grains underneath, which is almost twice the size of Ag grains grown directly on an atomically flat, inert substrate. Measurements of the critical thickness required for electrical conduction in the films verify this unanticipated growth behavior. We suggest that the larger Ag grains form because they tend to nucleate at the grain boundaries in the granular Pb film and consequently grow to sizes comparable to the underlying Pb grains.   

\begin{acknowledgments}
 We acknowledge the support of NSF-DMR980193 and NSF-DMR0203608. 
\end{acknowledgments}


\begin{thebibliography}{28}
\expandafter\ifx\csname natexlab\endcsname\relax\def\natexlab#1{#1}\fi
\expandafter\ifx\csname bibnamefont\endcsname\relax
  \def\bibnamefont#1{#1}\fi
\expandafter\ifx\csname bibfnamefont\endcsname\relax
  \def\bibfnamefont#1{#1}\fi
\expandafter\ifx\csname citenamefont\endcsname\relax
  \def\citenamefont#1{#1}\fi
\expandafter\ifx\csname url\endcsname\relax
  \def\url#1{\texttt{#1}}\fi
\expandafter\ifx\csname urlprefix\endcsname\relax\def\urlprefix{URL }\fi
\providecommand{\bibinfo}[2]{#2}
\providecommand{\eprint}[2][]{\url{#2}}

\bibitem[{\citenamefont{Buckel and Hilsch}(1954)}]{Buckel}
\bibinfo{author}{\bibfnamefont{W.}~\bibnamefont{Buckel}} \bibnamefont{and}
  \bibinfo{author}{\bibfnamefont{R.}~\bibnamefont{Hilsch}},
  \bibinfo{journal}{Z. Phys.} \textbf{\bibinfo{volume}{138}},
  \bibinfo{pages}{109} (\bibinfo{year}{1954}).

\bibitem[{\citenamefont{Buckel}(1954)}]{Buckel2}
\bibinfo{author}{\bibfnamefont{W.}~\bibnamefont{Buckel}}, \bibinfo{journal}{Z.
  Phys.} \textbf{\bibinfo{volume}{138}}, \bibinfo{pages}{136}
  (\bibinfo{year}{1954}).

\bibitem[{\citenamefont{Bulow and Buckel}(1956)}]{Bulow}
\bibinfo{author}{\bibfnamefont{H.}~\bibnamefont{Bulow}} \bibnamefont{and}
  \bibinfo{author}{\bibfnamefont{W.}~\bibnamefont{Buckel}},
  \bibinfo{journal}{Z. Phys.} \textbf{\bibinfo{volume}{145}},
  \bibinfo{pages}{141} (\bibinfo{year}{1956}).

\bibitem[{\citenamefont{Strongin et~al.}(1970)\citenamefont{Strongin, Thompson,
  Kammerer, and E.}}]{Strongin}
\bibinfo{author}{\bibfnamefont{M.}~\bibnamefont{Strongin}},
  \bibinfo{author}{\bibfnamefont{R.~S.} \bibnamefont{Thompson}},
  \bibinfo{author}{\bibfnamefont{O.~F.} \bibnamefont{Kammerer}},
  \bibnamefont{and} \bibinfo{author}{\bibfnamefont{C.~J.} \bibnamefont{E.}},
  \bibinfo{journal}{Phys. Rev. B} \textbf{\bibinfo{volume}{1}},
  \bibinfo{pages}{1078} (\bibinfo{year}{1970}).

\bibitem[{\citenamefont{Bergmann}(1984)}]{Bergmann}
\bibinfo{author}{\bibfnamefont{G.}~\bibnamefont{Bergmann}},
  \bibinfo{journal}{Phys. Rep.} \textbf{\bibinfo{volume}{107}},
  \bibinfo{pages}{1} (\bibinfo{year}{1984}).

\bibitem[{\citenamefont{Hsu and Valles}(1995)}]{Hsu}
\bibinfo{author}{\bibfnamefont{S.~Y.} \bibnamefont{Hsu}} \bibnamefont{and}
  \bibinfo{author}{\bibfnamefont{J.~M.} \bibnamefont{Valles}},
  \bibinfo{journal}{Phys. Rev. Lett.} \textbf{\bibinfo{volume}{74}},
  \bibinfo{pages}{2331} (\bibinfo{year}{1995}).

\bibitem[{\citenamefont{Liu et~al.}(1992)\citenamefont{Liu, Nease, Mcgreer, and
  Goldman}}]{Liu}
\bibinfo{author}{\bibfnamefont{Y.}~\bibnamefont{Liu}},
  \bibinfo{author}{\bibfnamefont{B.}~\bibnamefont{Nease}},
  \bibinfo{author}{\bibfnamefont{K.~A.} \bibnamefont{Mcgreer}},
  \bibnamefont{and} \bibinfo{author}{\bibfnamefont{A.~M.}
  \bibnamefont{Goldman}}, \bibinfo{journal}{Europhys. Lett.}
  \textbf{\bibinfo{volume}{19}}, \bibinfo{pages}{409} (\bibinfo{year}{1992}).

\bibitem[{\citenamefont{Dynes et~al.}(1978)\citenamefont{Dynes, Garno, and
  Rowell}}]{Dynes}
\bibinfo{author}{\bibfnamefont{R.~C.} \bibnamefont{Dynes}},
  \bibinfo{author}{\bibfnamefont{J.~P.} \bibnamefont{Garno}}, \bibnamefont{and}
  \bibinfo{author}{\bibfnamefont{J.~M.} \bibnamefont{Rowell}},
  \bibinfo{journal}{Phys. Rev. Lett.} \textbf{\bibinfo{volume}{40}},
  \bibinfo{pages}{479} (\bibinfo{year}{1978}).

\bibitem[{\citenamefont{Haviland et~al.}(1989)\citenamefont{Haviland, Liu, and
  Goldman}}]{Haviland}
\bibinfo{author}{\bibfnamefont{D.~B.} \bibnamefont{Haviland}},
  \bibinfo{author}{\bibfnamefont{Y.}~\bibnamefont{Liu}}, \bibnamefont{and}
  \bibinfo{author}{\bibfnamefont{A.~M.} \bibnamefont{Goldman}},
  \bibinfo{journal}{Phys. Rev. Lett.} \textbf{\bibinfo{volume}{62}},
  \bibinfo{pages}{2180} (\bibinfo{year}{1989}).

\bibitem[{\citenamefont{White et~al.}(1986)\citenamefont{White, Dynes, and
  Garno}}]{White}
\bibinfo{author}{\bibfnamefont{A.~E.} \bibnamefont{White}},
  \bibinfo{author}{\bibfnamefont{R.~C.} \bibnamefont{Dynes}}, \bibnamefont{and}
  \bibinfo{author}{\bibfnamefont{J.~P.} \bibnamefont{Garno}},
  \bibinfo{journal}{Phys. Rev. B} \textbf{\bibinfo{volume}{33}},
  \bibinfo{pages}{3549} (\bibinfo{year}{1986}).

\bibitem[{\citenamefont{Jaeger et~al.}(1986)\citenamefont{Jaeger, Haviland,
  Goldman, and Orr}}]{Jaeger}
\bibinfo{author}{\bibfnamefont{H.~M.} \bibnamefont{Jaeger}},
  \bibinfo{author}{\bibfnamefont{D.~B.} \bibnamefont{Haviland}},
  \bibinfo{author}{\bibfnamefont{A.~M.} \bibnamefont{Goldman}},
  \bibnamefont{and} \bibinfo{author}{\bibfnamefont{B.~G.} \bibnamefont{Orr}},
  \bibinfo{journal}{Phys. Rev. B} \textbf{\bibinfo{volume}{34}},
  \bibinfo{pages}{4920} (\bibinfo{year}{1986}).

\bibitem[{\citenamefont{Barber and Glover}(1990)}]{Barber}
\bibinfo{author}{\bibfnamefont{R.~P.} \bibnamefont{Barber}, \bibfnamefont{Jr.}}
  \bibnamefont{and} \bibinfo{author}{\bibfnamefont{R.~E.} \bibnamefont{Glover},
  \bibfnamefont{III}}, \bibinfo{journal}{Phys. Rev. B}
  \textbf{\bibinfo{volume}{42}}, \bibinfo{pages}{6754} (\bibinfo{year}{1990}).

\bibitem[{\citenamefont{Markiewicz et~al.}(1988)\citenamefont{Markiewicz,
  Shiffman, and Ho}}]{Markiewicz}
\bibinfo{author}{\bibfnamefont{R.~S.} \bibnamefont{Markiewicz}},
  \bibinfo{author}{\bibfnamefont{C.~A.} \bibnamefont{Shiffman}},
  \bibnamefont{and} \bibinfo{author}{\bibfnamefont{W.}~\bibnamefont{Ho}},
  \bibinfo{journal}{J. Low Temp. Phys.} \textbf{\bibinfo{volume}{71}},
  \bibinfo{pages}{175} (\bibinfo{year}{1988}).

\bibitem[{\citenamefont{Danilov et~al.}(1996)\citenamefont{Danilov, Kubatkin,
  Landau, and Rinderer}}]{Danilov}
\bibinfo{author}{\bibfnamefont{A.~V.} \bibnamefont{Danilov}},
  \bibinfo{author}{\bibfnamefont{S.~E.} \bibnamefont{Kubatkin}},
  \bibinfo{author}{\bibfnamefont{I.~L.} \bibnamefont{Landau}},
  \bibnamefont{and} \bibinfo{author}{\bibfnamefont{L.}~\bibnamefont{Rinderer}},
  \bibinfo{journal}{J. Low Temp. Phys.} \textbf{\bibinfo{volume}{103}},
  \bibinfo{pages}{35} (\bibinfo{year}{1996}).

\bibitem[{\citenamefont{Danilov et~al.}(1995)\citenamefont{Danilov, Kubatkin,
  Landau, Parshin, and Rinderer}}]{Danilov2}
\bibinfo{author}{\bibfnamefont{A.~V.} \bibnamefont{Danilov}},
  \bibinfo{author}{\bibfnamefont{S.~E.} \bibnamefont{Kubatkin}},
  \bibinfo{author}{\bibfnamefont{I.~L.} \bibnamefont{Landau}},
  \bibinfo{author}{\bibfnamefont{I.~A.} \bibnamefont{Parshin}},
  \bibnamefont{and} \bibinfo{author}{\bibfnamefont{L.}~\bibnamefont{Rinderer}},
  \bibinfo{journal}{Phys. Rev. B} \textbf{\bibinfo{volume}{51}},
  \bibinfo{pages}{5514} (\bibinfo{year}{1995}).

\bibitem[{\citenamefont{Ekinci and Valles~Jr.}(1998)}]{Ekinci1998}
\bibinfo{author}{\bibfnamefont{K.~L.} \bibnamefont{Ekinci}} \bibnamefont{and}
  \bibinfo{author}{\bibfnamefont{J.~M.} \bibnamefont{Valles~Jr.}},
  \bibinfo{journal}{Phys. Rev. B} \textbf{\bibinfo{volume}{58}},
  \bibinfo{pages}{7347} (\bibinfo{year}{1998}).

\bibitem[{\citenamefont{Ekinci and Valles~Jr.}(1999)}]{Ekinci1999}
\bibinfo{author}{\bibfnamefont{K.~L.} \bibnamefont{Ekinci}} \bibnamefont{and}
  \bibinfo{author}{\bibfnamefont{J.~M.} \bibnamefont{Valles~Jr.}},
  \bibinfo{journal}{Phys. Rev. Lett.} \textbf{\bibinfo{volume}{82}},
  \bibinfo{pages}{1518} (\bibinfo{year}{1999}).

\bibitem[{\citenamefont{Merchant et~al.}(2001)\citenamefont{Merchant, Ostrick,
  Barber, and Dynes}}]{Merchant}
\bibinfo{author}{\bibfnamefont{L.}~\bibnamefont{Merchant}},
  \bibinfo{author}{\bibfnamefont{J.}~\bibnamefont{Ostrick}},
  \bibinfo{author}{\bibfnamefont{R.~P.} \bibnamefont{Barber}},
  \bibnamefont{and} \bibinfo{author}{\bibfnamefont{R.~C.} \bibnamefont{Dynes}},
  \bibinfo{journal}{Phys. Rev. B} \textbf{\bibinfo{volume}{63}},
  \bibinfo{pages}{134508} (\bibinfo{year}{2001}).

\bibitem[{\citenamefont{Kouh and Valles}(2003)}]{Kouh}
\bibinfo{author}{\bibfnamefont{T.}~\bibnamefont{Kouh}} \bibnamefont{and}
  \bibinfo{author}{\bibfnamefont{J.~M.} \bibnamefont{Valles}},
  \bibinfo{journal}{Phys. Rev. B} \textbf{\bibinfo{volume}{67}},
  \bibinfo{pages}{140506} (\bibinfo{year}{2003}).

\bibitem[{\citenamefont{Long et~al.}(2004)\citenamefont{Long, Stewart~Jr.,
  Kouh, and Valles~Jr.}}]{Long}
\bibinfo{author}{\bibfnamefont{Z.}~\bibnamefont{Long}},
  \bibinfo{author}{\bibfnamefont{M.~D.} \bibnamefont{Stewart~Jr.}},
  \bibinfo{author}{\bibfnamefont{T.}~\bibnamefont{Kouh}}, \bibnamefont{and}
  \bibinfo{author}{\bibfnamefont{J.~M.} \bibnamefont{Valles~Jr.}},
  \bibinfo{journal}{arXiv:cond-mat/0409056}  (\bibinfo{year}{2004}).

\bibitem[{\citenamefont{Ekinci and Valles~Jr.}(1997)}]{Ekinci1997}
\bibinfo{author}{\bibfnamefont{K.~L.} \bibnamefont{Ekinci}} \bibnamefont{and}
  \bibinfo{author}{\bibfnamefont{J.~M.} \bibnamefont{Valles~Jr.}},
  \bibinfo{journal}{Rev. Sci. Instrum.} \textbf{\bibinfo{volume}{68}},
  \bibinfo{pages}{4152} (\bibinfo{year}{1997}).

\bibitem[{\citenamefont{Ekinci}(1999)}]{EkinciThesis}
\bibinfo{author}{\bibfnamefont{K.~L.} \bibnamefont{Ekinci}},
  \bibinfo{type}{Ph.d. thesis}, \bibinfo{school}{Brown University}
  (\bibinfo{year}{1999}).

\bibitem[{\citenamefont{Ekinci and Valles}(1998)}]{Ekinci19982}
\bibinfo{author}{\bibfnamefont{K.~L.} \bibnamefont{Ekinci}} \bibnamefont{and}
  \bibinfo{author}{\bibfnamefont{J.~M.} \bibnamefont{Valles}},
  \bibinfo{journal}{Acta Materialia} \textbf{\bibinfo{volume}{46}},
  \bibinfo{pages}{4549} (\bibinfo{year}{1998}).

\bibitem[{\citenamefont{Hu et~al.}(2003)\citenamefont{Hu, Noda, and
  Komiyama}}]{Hu}
\bibinfo{author}{\bibfnamefont{M.~H.} \bibnamefont{Hu}},
  \bibinfo{author}{\bibfnamefont{S.}~\bibnamefont{Noda}}, \bibnamefont{and}
  \bibinfo{author}{\bibfnamefont{H.}~\bibnamefont{Komiyama}},
  \bibinfo{journal}{J. Appl. Phys.} \textbf{\bibinfo{volume}{93}},
  \bibinfo{pages}{9336} (\bibinfo{year}{2003}).

\bibitem[{\citenamefont{Egelhoff and Jacob}(1989)}]{Egelhoff}
\bibinfo{author}{\bibfnamefont{W.~F.} \bibnamefont{Egelhoff}} \bibnamefont{and}
  \bibinfo{author}{\bibfnamefont{I.}~\bibnamefont{Jacob}},
  \bibinfo{journal}{Phys. Rev. Lett.} \textbf{\bibinfo{volume}{62}},
  \bibinfo{pages}{921} (\bibinfo{year}{1989}).

\bibitem[{\citenamefont{Ferris et~al.}(2000)\citenamefont{Ferris, Kushmerick,
  Johnson, and Weiss}}]{Ferris}
\bibinfo{author}{\bibfnamefont{J.~H.} \bibnamefont{Ferris}},
  \bibinfo{author}{\bibfnamefont{J.~G.} \bibnamefont{Kushmerick}},
  \bibinfo{author}{\bibfnamefont{J.~A.} \bibnamefont{Johnson}},
  \bibnamefont{and} \bibinfo{author}{\bibfnamefont{P.~S.} \bibnamefont{Weiss}},
  \bibinfo{journal}{Surf. Sci.} \textbf{\bibinfo{volume}{446}},
  \bibinfo{pages}{112} (\bibinfo{year}{2000}).

\bibitem[{\citenamefont{Evans et~al.}(1990)\citenamefont{Evans, Sanders, Thiel,
  and Depristo}}]{Evans}
\bibinfo{author}{\bibfnamefont{J.~W.} \bibnamefont{Evans}},
  \bibinfo{author}{\bibfnamefont{D.~E.} \bibnamefont{Sanders}},
  \bibinfo{author}{\bibfnamefont{P.~A.} \bibnamefont{Thiel}}, \bibnamefont{and}
  \bibinfo{author}{\bibfnamefont{A.~E.} \bibnamefont{Depristo}},
  \bibinfo{journal}{Phys. Rev. B} \textbf{\bibinfo{volume}{41}},
  \bibinfo{pages}{5410} (\bibinfo{year}{1990}).

\bibitem[{\citenamefont{Stoldt et~al.}(2000)\citenamefont{Stoldt, Caspersen,
  Bartelt, Jenks, Evans, and Thiel}}]{Stoldt}
\bibinfo{author}{\bibfnamefont{C.~R.} \bibnamefont{Stoldt}},
  \bibinfo{author}{\bibfnamefont{K.~J.} \bibnamefont{Caspersen}},
  \bibinfo{author}{\bibfnamefont{M.~C.} \bibnamefont{Bartelt}},
  \bibinfo{author}{\bibfnamefont{C.~J.} \bibnamefont{Jenks}},
  \bibinfo{author}{\bibfnamefont{J.~W.} \bibnamefont{Evans}}, \bibnamefont{and}
  \bibinfo{author}{\bibfnamefont{P.~A.} \bibnamefont{Thiel}},
  \bibinfo{journal}{Phys. Rev. Lett.} \textbf{\bibinfo{volume}{85}},
  \bibinfo{pages}{800} (\bibinfo{year}{2000}).

\end{thebibliography}

\end{document}